# Imaging Atomic-scale Effects of High Energy Ion Irradiation on Superconductivity and Vortex Pinning in Fe(Se,Te)

F. Massee[1,2,3,†], P. O. Sprau[1,2,†], Y. L. Wang[4], J. C. Davis[1,2,5,6], G. Ghigo[7,8], G. Gu[1], and W. K. Kwok[4]

[1] CMPMS Department, Brookhaven National Laboratory, Upton, NY 11973, USA
[2] LASSP, Department of Physics, Cornell University, Ithaca, NY 14853, USA.
[3] Laboratoire de Physique des Solides, Universite Paris-Sud, Orsay, France
[4] Materials Science Division, Argonne National Laboratory, Argonne, Illinois 60439, USA
[5] School of Physics and Astronomy, University of St Andrews, St Andrews, Fife KY16 9SS, UK.
[6] Kavli Institute at Cornell for Nanoscale Science, Cornell University, Ithaca, NY 14853, USA.
[7] Department of Applied Science and Technology, Politecnico di Torino, 10129 Torino, ITALY
[8] Istituto Nazionale di Fisica Nucleare, Sez. Torino, 10125 Torino, ITALY
† These authors contributed equally to this work

**Maximizing the sustainable supercurrent density, $J_C$, is crucial to high current applications of superconductivity and, to achieve this, preventing dissipative motion of quantized vortices is key. Irradiation of superconductors with high-energy heavy ions can be used to create nanoscale defects that act as deep pinning potentials for vortices. This approach holds unique promise for high current applications of iron-based superconductors because $J_c$ amplification persists to much higher radiation doses than in cuprate superconductors without significantly altering the superconducting critical temperature. However, for these compounds virtually nothing is known about the atomic scale interplay of the crystal damage from the high-energy ions, the superconducting order parameter, and the vortex pinning processes. Here, we visualize the atomic-scale effects of irradiating FeSe$_x$Te$_{1-x}$ with 249 MeV Au ions and find two distinct effects: compact nanometer-sized regions of crystal disruption or 'columnar defects', plus a higher density of single atomic-site 'point' defects probably from secondary scattering. We show directly that the superconducting order is virtually annihilated within the former while suppressed by the latter.**

**Simultaneous atomically-resolved images of the columnar crystal defects, the superconductivity, and the vortex configurations, then reveal how a mixed pinning landscape is created, with the strongest pinning occurring at metallic-core columnar defects and secondary pinning at clusters of point-like defects, followed by collective pinning at higher fields.**

SUMMARY: By introducing techniques to visualize atomic-scale effects of high energy ion irradiation, we achieve simultaneous atomically-resolved images of the columnar crystal defects, the superconductivity, and the vortex configurations, revealing the complexity of the pinning landscape where the strongest pinning occurs at the ion-track sites and secondary pinning at clusters of point-like defects.

Iron-based superconductors[1] are promising for high $J_C$ applications[2] because of a nexus of several materials characteristics[3]. First, the maximum critical field $H_{C2}$ is very high at low temperatures[4,5] while the compounds also exhibit rather isotropic superconductivity. Second, as in the cuprates[6], $J_C$ can be strongly enhanced by high energy ion irradiation[2,7]. Finally, the irradiation leaves $T_C$ virtually unchanged to a degree unknown in cuprate high temperature superconductors. Therefore, if engineered control of $J_C$ could be achieved under these circumstances, these materials could be very favorable for high-current/high-field applications. The theoretical understanding necessary for such materials engineering requires specific atomic-scale knowledge, including the structure of ion-induced columnar defects, along with their local influence on the superconductivity. For example, detailed knowledge of a columnar defect's internal conductivity and of its size with respect to the superconducting coherence length are required to predict quantitatively its interaction with a vortex core[8,9]. Imaging of high-energy ion induced columnar defects has been achieved using transmission electron microscopy[6,11-14], and visualization of irradiation induced disordered vortex configurations[15,16] have been achieved by STM. However, to our knowledge simultaneous atomic-scale visualization of the effects of high-energy ions on the crystal, their impact on the superconductivity, plus the resulting responses of the pinned-vortex configurations, has not been achieved for any type of superconductor.

To initiate such studies, we choose $FeSe_xTe_{1-x}$ [17]. In bulk single crystal form, its transition temperature can reach up to ~15 K with $H_{C2}$ at tens of Tesla[18]; in thin films, critical fields are enhanced and $T_C \sim 100$ K has been reported for unit-cell-thick monolayers of FeSe[19]. Here we use a [3]He-refrigerator-based spectroscopic imaging scanning tunneling microscope[20] (SI-STM) into which the $FeSe_xTe_{1-x}$ samples are inserted to be cleaved in cryogenic ultra-high vacuum at T<20 K. This technique consists of making atomically resolved and registered images of the surface topography T($r$) simultaneously with tip-sample differential tunneling conductance images $g(\mathbf{r},E=eV) \equiv dI/dV(\mathbf{r},E=eV)$ measured as a function of both location $\mathbf{r}$ and electron energy $E$. Figure 1a shows a typical T($\mathbf{r}$) of the TeSe termination layer with individual Te/Se atomic sites clearly visible[21,22]. In the superconducting phase at T=0.25 K, our measured $g(\mathbf{r},E)$ spectra are then fully gapped with clear coherence peaks[23] (arrows Fig. 1b) and a spatially homogeneous superconducting gap $\Delta$ (Fig. 1b). Upon application of magnetic field, the vortex lattice is observed in the zero bias conductance map, as shown in Fig. 1c and in the autocorrelation function, depicted in Fig. 1d, we observe a hexagonal pattern, pointing to a real space vortex lattice that remains overall with hexagonal order.

Single crystals of $FeSe_{0.45}Te_{0.55}$ from the same batch as in Fig. 1 were then irradiated with 249 MeV Au ions using a fluence of $1.93 \cdot 10^{15}$ m$^{-2}$ so that the 'dose equivalent field' is $B_\phi$= 4 Tesla ( this is the field ideally corresponding to a fluxon per incident ion ). However, few-hundreds-MeV heavy ions in metallic Fe-based superconductors create defect tracks that are expected to be discontinuous, thus the actual columnar defect density may be lower than the fluence[12]. Figure 2a shows a high resolution T($\mathbf{r}$) typical of the irradiated $FeSe_xTe_{1-x}$ crystals, in which two striking new features are apparent. The first consists of large (radius $\sim$ 1.5 nm) amorphous regions (e.g. red circles in Fig. 2a) with a surface coverage equivalent to a matching field of about 2 Tesla, in this field of view (FOV). The second type of feature occurs in larger numbers, and consists of an atomic-scale point defect (e.g. blue circles) centered in between Se/Te sites, i.e. at the Fe site in the layer below the surface. These are reminiscent of excess iron atoms observed in other studies[21,24]. Since on multiple pristine samples from the same growth batch we never see such excess Fe atoms, we speculate that the heavy ion irradiation has displaced Fe atoms into these sites. The matrix in which these two ion-induced defect types are detected exhibits an unperturbed $FeSe_{0.45}Te_{0.55}$ termination-layer morphology (inset Fig. 2a). See Supplementary Information section 2 for additional, more detailed, studies of these two types of ion-induced crystal defect.

Atomic-scale imaging reveals that columnar defects exhibit an amorphous crystal structure in a region with a diameter of approximately 3 nm. For each, the impact on superconductivity is its annihilation, as shown in Fig. 2b which compares the average g(E) spectrum (grey) to that at the center of ion-induced columnar defects (red). These data demonstrate directly that the columnar defect cores of Fe(Se,Te) are metallic. By contrast, the signature of superconductivity in each point-defect spectrum (blue Fig. 2c) is suppressed significantly relative to the average g(E) spectrum (grey Fig. 2c) meaning that these regions should individually provide weaker pinning sites. A short discussion of the properties of the defects far beyond the energy scale of superconductivity is available in the supplementary information. From a global perspective, the effects on the superconductivity of the high-energy ion irradiation are both profound and somewhat unexpected. Specifically, the $g(\mathbf{r},E)$ images measured on irradiated samples are no longer characterized by a homogeneous, full superconducting gap, but show a strong spatial variation with a finite differential conductance at zero bias everywhere. To illustrate the effects on the superconductivity in the FOV of Fig. 2a, we define the normalized function

$$\mathsf{F}(\boldsymbol{r}) = \frac{g(\boldsymbol{r},\Delta) - g(\boldsymbol{r},0)}{\overline{g(\boldsymbol{r})}} \qquad (1)$$

as a measure of the strength of the spectral signature of superconductivity. Here $g(\boldsymbol{r},\Delta)$ is the sum of g($\boldsymbol{r}$,E) over the energy region of coherence peaks $\Delta \sim (\pm 1.5 \leq E \leq \pm 2.5\ meV)$, g($\boldsymbol{r}$,0) is the sum of g($\boldsymbol{r}$,E) over the energy window centered on zero $(-0.5 \leq E \leq 0.5\ meV)$, while the average in the denominator runs over $E = \pm 5 meV$. Then, for F>1, the superconducting peak-to-dip difference is at least as large as the approximate normal state absolute conductance, hence there is a well-defined superconducting spectral signature. For F<0.2, the superconducting signature is on the order of, or smaller than the noise level, meaning superconductivity is completely suppressed. The $F(\mathbf{r})$ image measured in the FOV of Fig. 2a is shown in Fig. 2d and reveals the atomic-scale spatial arrangements of damage to the superconductivity as a result of heavy-ion irradiation in Fe(Se,Te), see Supplementary Information section 3 for comparison with the identical analysis of the pristine sample. Less than 50% of this (and all equivalent) FOV is weakly affected by irradiation (dark blue). The three columnar defects each exhibit complete suppression of the superconductivity but only within a radius of about 1.5nm so that, in themselves, they could not impact the overall superconductivity to the degree observed. In fact, it is the combined effect of the more than 20 point defects that dominate, especially when several are clustered within a mutual radius of ~3 nm with a resulting strong suppression of superconductivity. Additional analysis on the relationship between point defect position and order parameter suppression is provided in the supplementary information. The further remarkable thing about this situation is that $T_c$ is barely suppressed: by less than 1 K, and Jc is strongly enhanced (see Supplementary Information section 1). To understand the microscopics of vortex pinning by this complex superconducting landscape is the objective.

The field dependence of the vortex distribution process in irradiated Fe(Se,Te) is next determined. In an identical FOV we measure a series of T=0.25 K electronic structure images g($\mathbf{r}$, $E$, $B$), where B is the magnitude of the magnetic field applied perpendicular to the crystal surface. The classic signature of a vortex core when observed by measuring g($\mathbf{r}$,E) is the suppression of coherence peaks surrounding $E \sim \pm \Delta$ and the increase in zero-bias conductance surrounding E~0. A reasonable and practical way to detect vortices is to image the function

$$S(\mathbf{r}) = \big(g(\mathbf{r}, 0, B) - g(\mathbf{r}, 0, 0)\big) -$$
$$(g(\mathbf{r}, \Delta_-, B) - g(\mathbf{r}, \Delta_-, 0) + g(\mathbf{r}, \Delta_+, B) - g(\mathbf{r}, \Delta_+, 0) )/2 \quad (2)$$

which combines spectral weight from both primary phenomena near the core. Surprisingly, however, as exemplified in Fig. 3, the signature of vortices in the presence of columnar defects is not of this simple form. In fact, for fields up to about 2 Tesla for this FOV, we hardly see this classic signature of the vortex cores being introduced at all. Instead, the most common observation is that a circular 'halo' is detected in $S(\boldsymbol{r})$, surrounding columnar defects identified from the local crystal damage in T($\boldsymbol{r}$). Figure 3a shows a small FOV with ~7 columnar defects while the simultaneous Fig. 3b shows high resolution $S(\boldsymbol{r})$ measured at B=2 T. A vortex is (collectively) pinned within the red box in both figures that does not contain a columnar defect, and its signature in S($\boldsymbol{r}$) is as expected. However, the vortex pinned at a columnar defect shown with the yellow box in both images has a very distinct signature comprising of a 'halo' in S($\boldsymbol{r}$) surrounding the columnar defect; this vortex 'halo' signature is found at many columnar defects whose average topographic signature is shown in Fig. 3d and whose average S($\boldsymbol{r}$) is shown in Fig. 3c. A comparison between 'halo' signature and observable vortices is presented in the supplementary information. The concept is that the 'halo' is the signature of a pinned fluxon, but one where the conventional vortex core spectrum cannot be detected as the fluxon resides on a location of suppressed superconductivity at zero magnetic field.

Once the applied field exceeds about 2 Tesla for this FOV, the additional vortex core locations become more easily observable, exhibiting a reasonable example of the classic signature. In terms of $S(\boldsymbol{r})$ this is expected to be a bright circularly symmetric region of high |S| and of radius near one coherence length, which is what is observed. Under these circumstances, the field dependence of the configuration of vortex core locations can be determined directly. Figure 4a shows a typical FOV for such B-dependence studies with Fig. 4b the ion-induced damage to the superconductivity as determined by measuring F($\boldsymbol{r}$) simultaneously with Fig. 4a. The evolution of vortex locations with field is revealed directly in the measured S($\boldsymbol{r}$,B) images as shown in Figs. 4c-h . Below the damage equivalent field of 2 T, the vortices are rarely detectable as a circular region in S($\boldsymbol{r}$) as explained above (Fig. 3). Figure 4i shows the normalized cross-correlation of F($\boldsymbol{r}$) with S($\boldsymbol{r}$,B) in red, revealing that while there is little relation below 2 T, a strong positive correlation appears between regions of superconductivity (high F($\boldsymbol{r}$)) and the vortex signature S($\boldsymbol{r}$,B) at higher fields. The related anti-correlation at high fields of the topograph T($r$) (Fig. 4a) and S($r$,B) is shown in blue, and occurs because the regions of unperturbed superconductivity occur where little damage is detected in T($r$).

The interplay between ion-induced crystal damage, the heterogeneous superconductivity, and pinning of vortices revealed by these studies can be summarized as in Fig. 5. Figure 5a shows typical T($r$) with ~8 columnar defects plus many point defects. Despite the negligible impact on the superconducting $T_c$, the local superconductivity as estimated using $F(r)$ can be greatly impacted (Fig. 5b). While this effect is pronounced at columnar defects, the point defects, when occurring at high density, also have strong effect on $F(r)$. Then, from the field-dependence studies (Fig. 5c), we conclude that vortices are first very strongly pinned to the metallic-core columnar defects where the spectral signature of superconductivity is eradicated. This number is fixed and represented in Fig. 5d by ~8 grey patches and in 5e by grey components of the columns. At higher fields, observation of a strongly disordered vortex lattice appearing between the columnar defect sites indicates additional vortex pinning by clusters of point defects represented by orange circles in Fig. 5d and orange components of the columns in Fig. 5e. Finally, at highest fields, vortices begin to populate the areas of undamaged superconductivity and therefore to be collectively pinned as shown as dark blue patches in Fig. 5d and similar color components of the columns in Fig. 5e. Thus, the evolution of vortex configurations in a single field of view (e.g. Fig. 4) can be understood as a sequence of these three pinning processes.

Overall our studies reveal that a picture of vortices localized at damage tracks of the size of the coherence length in an otherwise unaffected superconductor is oversimplified. Instead, superconductivity is affected on a much larger scale, and a mixed pinning landscape is obtained where the strongest pinning occurs at the columnar defect sites, that on average are of the size near the coherence length as evidenced by the vortex halos we observe around them, and secondary pinning at clusters of point-like defects. Our finding that the amorphous cores of the Fe(Se,Te) columnar defects are metallic is significant as such cores exhibit quite different pinning potentials compared to those that are insulating, due to the distinct influence of superconducting proximity effect[8,9]. The discovery of such a complex mixed pinning landscape in high energy ion irradiated Fe(Se,Te) is also important because such a situation suppresses detrimental 'double-kink' vortex-creep[25,26], enabling even higher values for $J_c$ than in a scenario with columnar defects alone. Moreover, the novel combination of techniques that we introduce for simultaneous visualization of defects and superconductivity and vortex configurations, can greatly aid in predictive engineering of vortex matter in high temperature superconductors. This is because, in future, such measured spatial shapes and high-energy resolution spectral fingerprints of both vortices and heavy-ion induced defects, in combination with realistic multi-band Bogoliubov-deGennes theory[27] representing the identical real electronic environment, will be able to yield quantitative microscopic input parameters for massive Ginzburg-Landau type simulation of optimal vortex pinning. Finally, by using this same approach one could also pursue similar objectives in other materials such as cuprate high-Tc superconductors.

## Materials and Methods

High quality $FeSe_{0.45}Te_{0.55}$ single crystals were grown at Brookhaven National Laboratory. The samples were irradiated at room temperature at the Laboratori Nazionali di Legnaro of the Istituto Nazionale di Fisica Nucleare (INFN), Italy, using 249 MeV $Au^{17+}$ ions with a fluence $N = 1.93 \cdot 10^{15} \mathrm{m}^{-2}$. The beam current was 0.2 nA with a 0.56 $cm^2$ spot.[28] Magnetization measurements of both pristine and irradiated samples prior to insertion into the STM show a sharp transition with $T_c$ = 14 K ± 0.5 K, see Supplementary Information section 1 for more details. The samples were mechanically cleaved in cryogenic ultrahigh vacuum at $T{\sim}20$ K and directly inserted into the STM head at 4.2 K. Etched atomically sharp and stable tungsten tips with energy independent density of states were used.

Differential conductance measurements throughout used a standard lock-in amplifier. All topographic data shown was taken at E=-50 mV and I=50 pA.

**Acknowledgements:** Experimental studies are supported by the Center for Emergent Superconductivity, an Energy Frontier Research Center, headquartered at Brookhaven National Laboratory and funded by the U.S. Department of Energy, under DE-2009-BNL-PM015. Irradiations were performed in the framework of the INFN-Politecnico di Torino M.E.S.H. experiment.

F. Massee and P. O. Sprau performed STM measurements and data analysis. Y. L. Wang performed the magnetization and $J_c$ measurements and G. Ghigo performed irradiation with Au-ions. G. Gu synthesized Fe(Se, Te) samples. J. C. Davis and W. K. Kwok designed and supervised project.

We declare that no competing financial interests exist.

**Fig. 1**

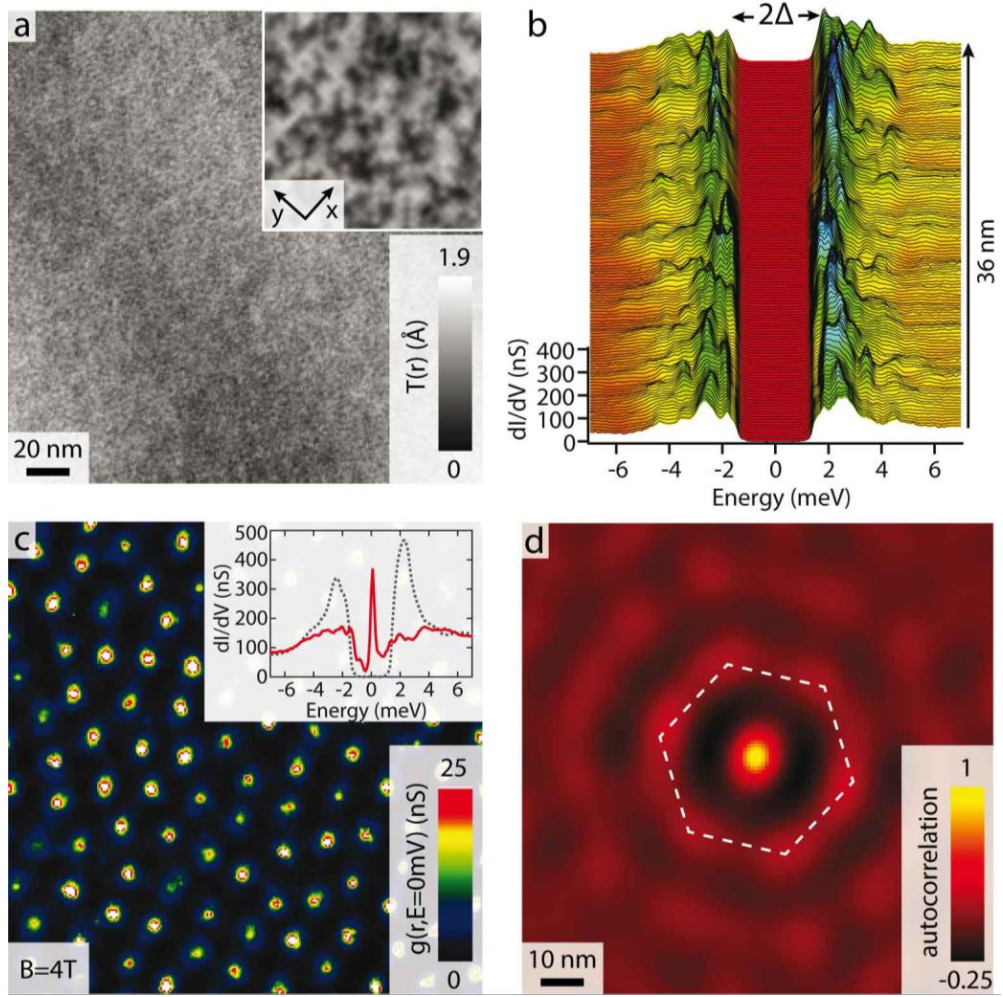

**Fig. 1  Visualizing Superconductivity in Pristine Fe(Se,Te)**

**a**     Large, atomically resolved constant current topograph T($r$) of FeSe$_{0.45}$Te$_{0.55}$.The inset shows an enlargement of the atomic lattice using the same color scale.

**b**     Differential conduction spectra at 270 mK taken along a line just outside the field of view shown in panel a: all spectra are fully gapped with clear coherence peaks (indicated by arrows). The multitude of peaks outside the gap reflect the multiband nature of the system.

**c**     Vortex lattice of pristine Fe(Se,Te). The lattice is predominantly hexagonal with only minor distortions due to native pinning of vortices. The inset shows the field of view average spectrum away from vortices (dashed) and a typical spectrum taken at the core of a vortex (red) at 270 mK.

**d**     Autocorrelation of (c) exemplifying the predominance of the hexagonal vortex structure.

**Fig. 2**

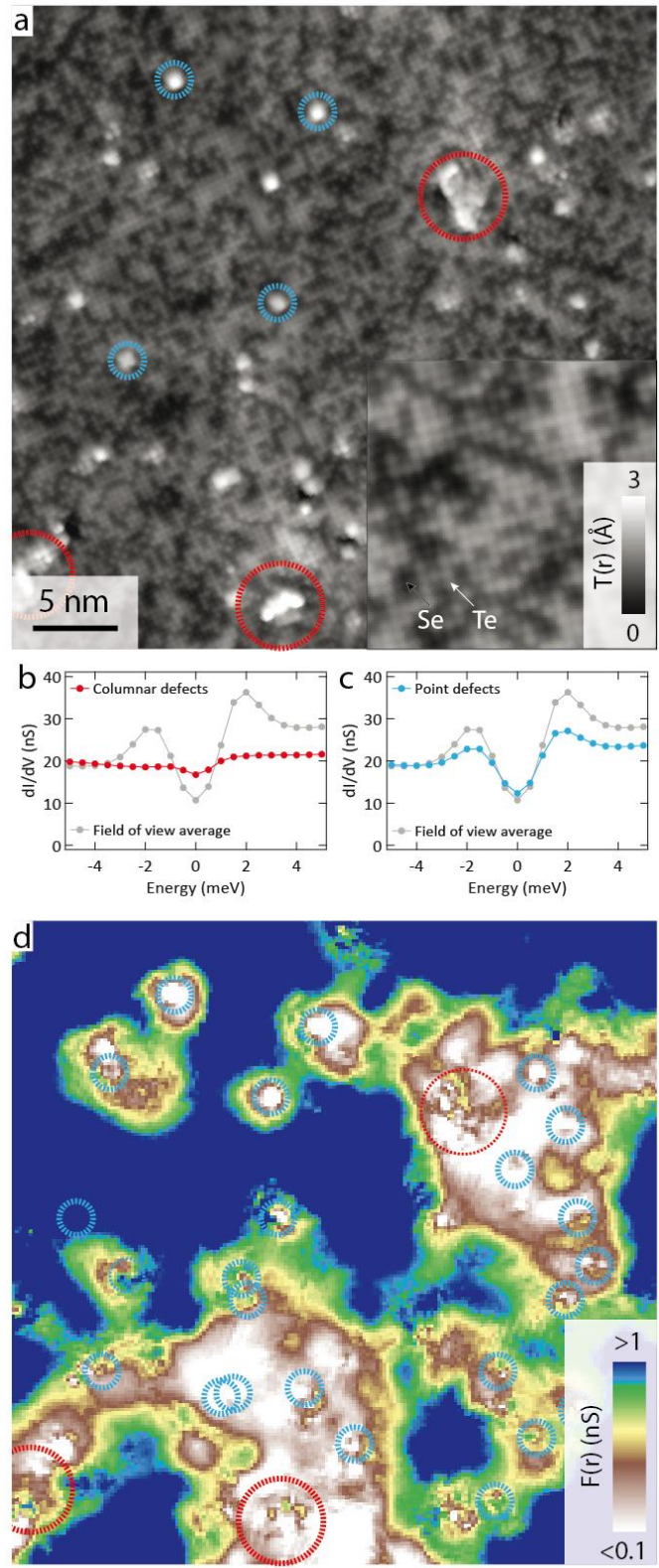

**Fig. 2   Impact on Fe(Se,Te) superconductivity of heavy ion-irradiation**

**a**    High resolution T($r$) of heavy ion irradiated FeSe$_{0.45}$Te$_{0.55}$. As for the pristine sample, the predominant feature is the binary Se/Te surface appearance (inset). Red and blue circles indicate the columnar and point defects that are both only observed after irradiation. Depending on the FOV, the observed damage track density during our SI-STM studies varies between 2T and 4T effective dose. This may be because the columnar defect tracks are discontinuous or because the distribution is sufficiently heterogeneous that FOV may not be large enough for an accurate statistical count.

**b**    Average differential conduction spectrum at 1.2 K of columnar defects in **a**: The superconducting signature in the tunnel spectrum is completely suppressed.

**c**    Average differential conduction spectrum at 1.2 K of point defects in **a**: The superconducting signature in the tunnel spectrum shows significant reduction.

**d**    F(r) as defined in text for the same field of view as depicted in **a**: Note the excellent correlation between suppressed superconductivity and position of columnar and clusters of point defects, marked by red and blue dashed circles, respectively.

**Fig. 3**

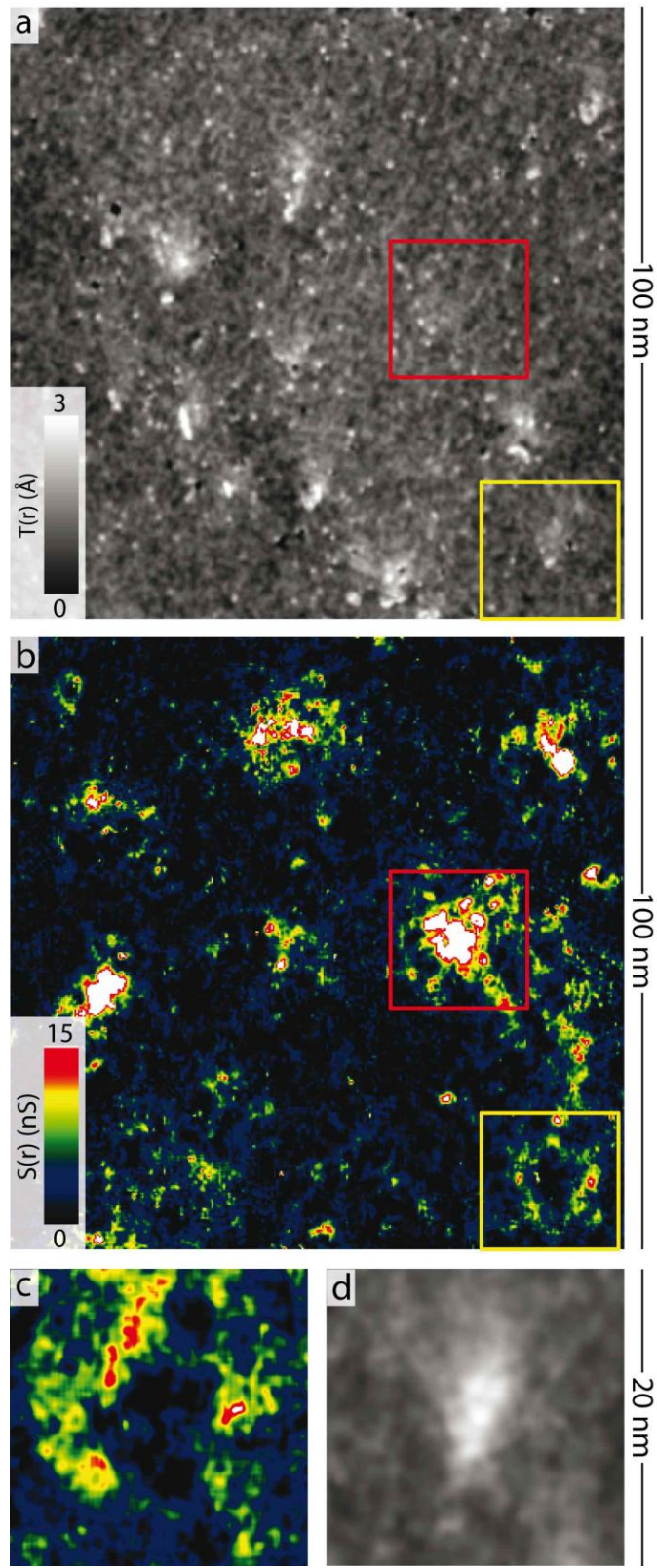

**Fig. 3   Vortex 'Halo' Surrounding Columnar Defects**

**a**      Constant current image T(r): The red and yellow rectangle mark the position of vortices shown in **b**.

**b**      S($r$) at 2 T of same field of view as in **a**: The red rectangle marks a vortex that is far away from columnar defects, while the yellow rectangle encircles the "halo" of a vortex pinned to the columnar defect marked by the yellow rectangle in **a**.

**c**      Average S($r$) of all columnar defects in **a**: Note the vortex halo surrounding a dark core due to the strong pinning of vortices to columnar defects at low fields.

**d**      Average T($r$) of all columnar defects in **a**: The average columnar defect position agrees very well with the location of the pinning center deduced from the average vortex halo shown in **c**.

**Fig. 4**

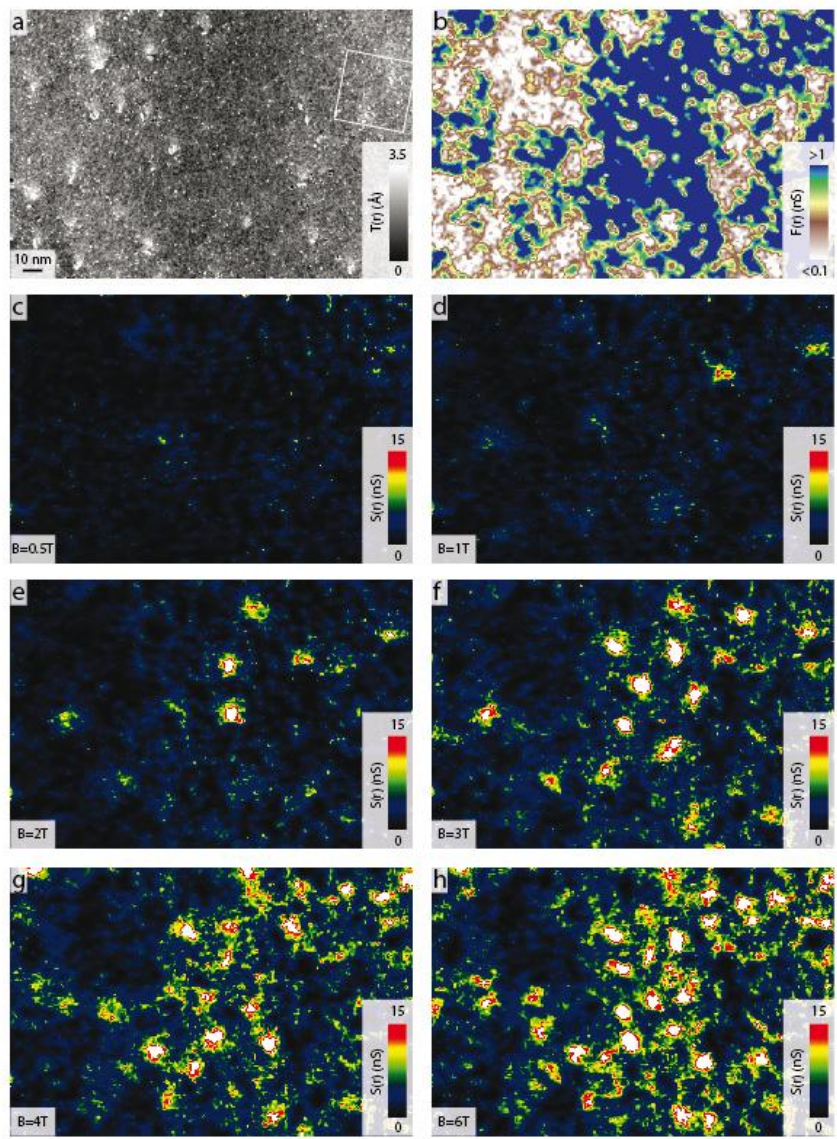

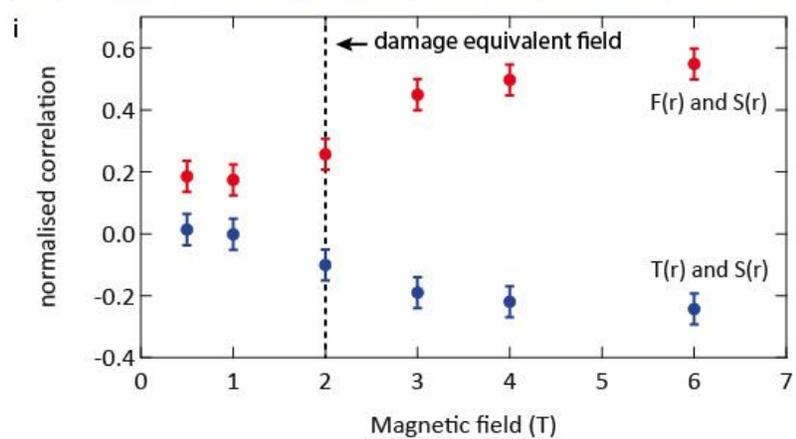

**Fig. 4 Evolution of Vortex Configurations with Magnetic Field**

**a**    Constant current image T(r) of same field of view studied in b-h.

**b**    F(r) of same field of view as in **a**.

**c-h**  S($\mathbf{r}$,*B)* for B=0.5, 1, 2, 3, 4 and 6 T respectively.

**i**    Normalized cross-correlation between S($\boldsymbol{r}$) and T($\boldsymbol{r}$) (blue), and S($\boldsymbol{r}$) and F($\boldsymbol{r}$)

(red), as a function of field.

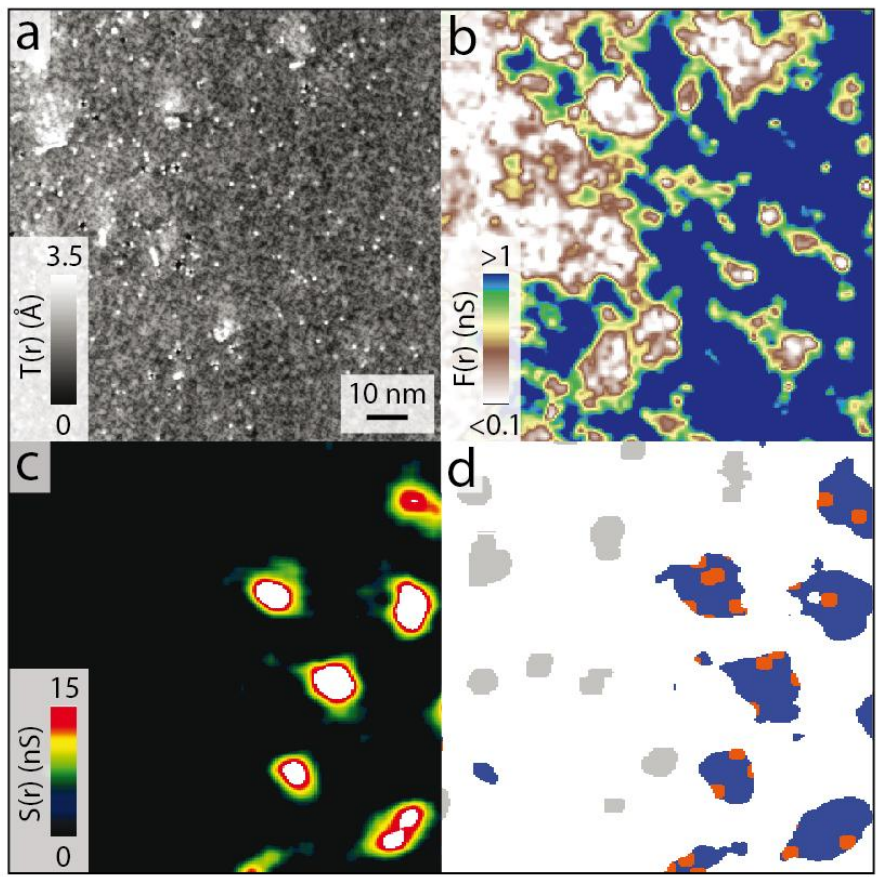

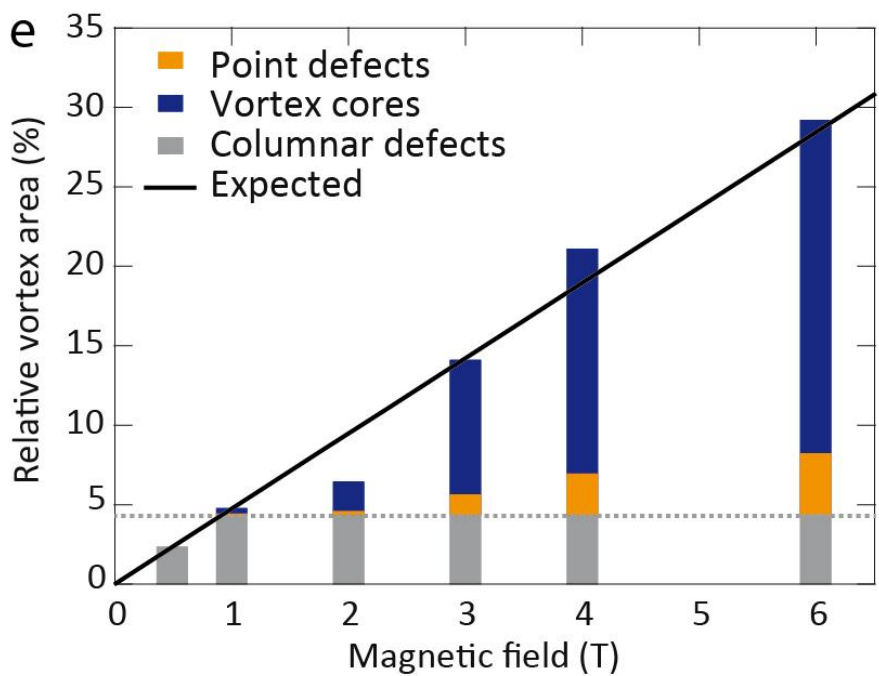

**Fig. 5 Overview of Vortex Pinning Sequence**

**a**      Constant current, T$(r)$, image taken on irradiated Fe(Se,Te), showing the columnar defects and point defects clearly.

**b**      F$(r)$ in the same field of view as **a**, illustrating the effect of the two types of defect on the superconducting tunneling signature.

**c**      Vortex image, S$(r)$, taken at 3 T on the area of panel **a**.

**d**      Color-coded breakdown of the interplay of irradiation induced defects, suppression of the spectral signature of superconductivity and vortex pinning. Note the overlap between vortices (blue) far away from columnar defects (grey) and point defects (orange). The vortex density is the area in percentage of the full field of view covered by the various features.

**e**      Histogram representing the relation between vortices, columnar and point defects distributed in a random, mixed pinning landscape created by swift ion irradiation. Data in the histogram was obtained by analysis of the whole field of view depicted in Fig. 4.

*Supplementary Information:*

**Imaging Atomic-scale Effects of High Energy Ion Irradiation on Superconductivity and Vortex Pinning in Fe(Se,Te)**

F. Massee[†], P. O. Sprau[†], Y. L. Wang, J. C. Davis, G. Ghigo, G. Gu,
and W. K. Kwok

## (I)    Sample characterization

In order to characterize the effect of irradiation with 249 MeV Au ions on the macroscopic properties of superconductivity in Fe(Se, Te) we measure the magnetization as a function of temperature and field. The limited penetration depth of the Au ions allows us to extract irradiated and pristine regions from the same sample measured by STM. For this purpose we remove several μm thick layers from opposite sides of the sample. A schematic drawing of the crystal with an estimated thickness for the irradiated and the pristine layers removed is presented in Fig. S1a.

Figure S1b presents the magnetization measurement and its derivative with respect to temperature as a function of temperature. $T_C$ is determined by the peak position in the derivative. $T_C$ remains practically unchanged after irradiation. Measured magnetization as a function of applied magnetic field is depicted in Fig. S1c. The enlarged magnetization loop of the irradiated sample indicates enhanced pinning by irradiation.

The critical current density $J_C$ is shown in Fig S1d. $J_C$ is enhanced after irradiation at all the measured magnetic fields up to 7 Tesla. The $J_C$ enhancement may be underestimated either due to a conservative estimate of the sample thickness for the pristine crystal (> 45 µm) and irradiated (< 10 µm) layers or due to some non-irradiated layers under the cleaved-off sample layers, as we are not sure how many layers were removed during STM sample cleaving.

## (II)   High Energy Ion Damage Characteristics

In this section we show the two types of damage in more detail. Figure S2a shows the same constant current image as in Fig. 3a of the main text, indicating all columnar defects with a red circle. There is a small uncertainty in the exact number of columnar defects as it is not always clear whether a defect is a single columnar defect or a cluster of point defects.

Shown in Fig. S2b are four enlargements of columnar defects. At the core of the columnar defect, no atomic lattice can be discerned; instead, an amorphous region with a larger corrugation is seen. In order to illustrate this point in more detail, we show the analysis of a columnar defect in real space and Fourier space in Fig. S3. Figure S3a shows a high resolution topograph around a columnar defect. In Figs. S3b and S3c we show isolated areas of the same size that contain the undisturbed atomic lattice and the columnar defect, respectively. There is no crystal lattice recognizable in the defect in real space. In order to substantiate the evidence for an amorphous core we take the Fourier transform of Figs. S3a to S3c which are shown in Figs. S3d to S3f. The Bragg peaks of the atomic lattice are clearly visible in Figs. S3d and S3e, but absent in Fig. S3f which instead is dominated by long wavelength signals. This is the expected result for the Fourier transform of a structure that lacks periodicity, as for example in form of the crystal lattice. Thus, there is strong evidence for an amorphous core in the columnar defects.

The point-like defects show up as bright spots on an otherwise unperturbed Se/Te surface, which are centered between the Se/Te atoms, strongly indicating that the defect is an excess iron atom on an interstitial iron site[1,2]. The complete absence of excess iron in our pristine samples leads us to conclude that the point defects we observe in the irradiated samples are irradiation induced excess iron sites.

## (III)   Order parameter suppression in the pristine sample

In this section we present the order parameter suppression in the pristine sample in comparison to the suppression observed after irradiation.

In Fig. S4a we present the topograph of the pristine sample showing no defects. The corresponding map of F(r) is depicted in Fig. S4b. There is no significant order parameter suppression in the whole field of view. This is in stark contrast to the irradiated sample. A topograph and F(r) map of the irradiated sample are shown in Figs. S4c and S4d. The size of the field of view is identical for both the irradiated and the pristine sample. Conductance maps used for the generation of F(r)-maps were recorded simultaneously with the topographs at 270 mK and 0 T in both cases.

In Fig. S4e we present the field of view average spectra of the pristine sample with two distinct areas in the irradiated sample: the area that is defect-free (F(r) > 1), and the area that contains the point and columnar defects (F(r) ≤1). Both regions show a decreased average gap size and an increase of density of states inside the superconducting gap, with the effect being more pronounced for the region containing the defects. Despite this general suppression of superconductivity in the differential conductance tunneling spectra, the bulk critical temperature did not change as shown in Fig. S1.

## (IV)  Effect of point defects on order parameter suppression

To illustrate the effect of the point-like defects on the order parameter seen with STM, we show a typical topograph and corresponding F(r) map in Fig. S5. The point-like defects are indicated with small black circles on the F(r) map. Clearly, there is a strong correlation between the location of the small defects and a suppressed order parameter. We note that there are a very small number of point-defects that do not seem to suppress superconductivity, in this work we focus on the majority that does suppress superconductivity though.

## (V)  Influence of irradiation induced defects on normal state

To investigate the effect of the irradiation induced defects on the LDOS in the energy range far beyond the superconducting gap we recorded a differential conductance map from -210 mV to +210 mV with a resolution of 15 mV at 1.2K. Point and columnar defects are clearly visible in the differential conductance layer at -30 mV, Fig. S6a, and the topograph in the same field of view, Fig. S6b. The average spectra for the point and columnar defects differ strongly from the average spectra in the defect free region, as can be seen in Fig. S6c. In the energy range between $\sim$ -100 meV and $\sim$ +100 meV the LDOS is enhanced at defect sites compared to the defect free regions, while below $\sim$ -100 meV and above $\sim$ +100 meV the LDOS at the defect sites is diminished. Around approximately +90 meV a broad resonance is observed for the defects which is absent in the undamaged region of the sample. The origin for both the resonance and the redistribution of LDOS to energies closer to the chemical potential is unknown at present. In general we conclude that the spectra display metallic character, possibly modified by the presence of strong scattering centers in comparison to the defect free regions.

## (VI)   Vortex 'halos'

In this section we illustrate that the 'halos' we find in S(r) at the location of columnar defects are indeed halos of vortices. Figure S7a reproduces the average S(r) at the columnar defect locations (main text Fig. 3c) showing the vortex 'halo'. In contrast, Fig. S7b shows the average of the four 'traditional' vortices that can be seen in the field of view of Fig. 3b. From this comparison it is evident that the hole inside the halo corresponds very well to the core of the average vortex, marked with a light blue circle in both images. In combination with the strong correlation of the halo signal to the columnar defects we infer that the halo is not a collection of points but indeed the halo of a vortex pinned to a columnar defect.

**Fig. S1**

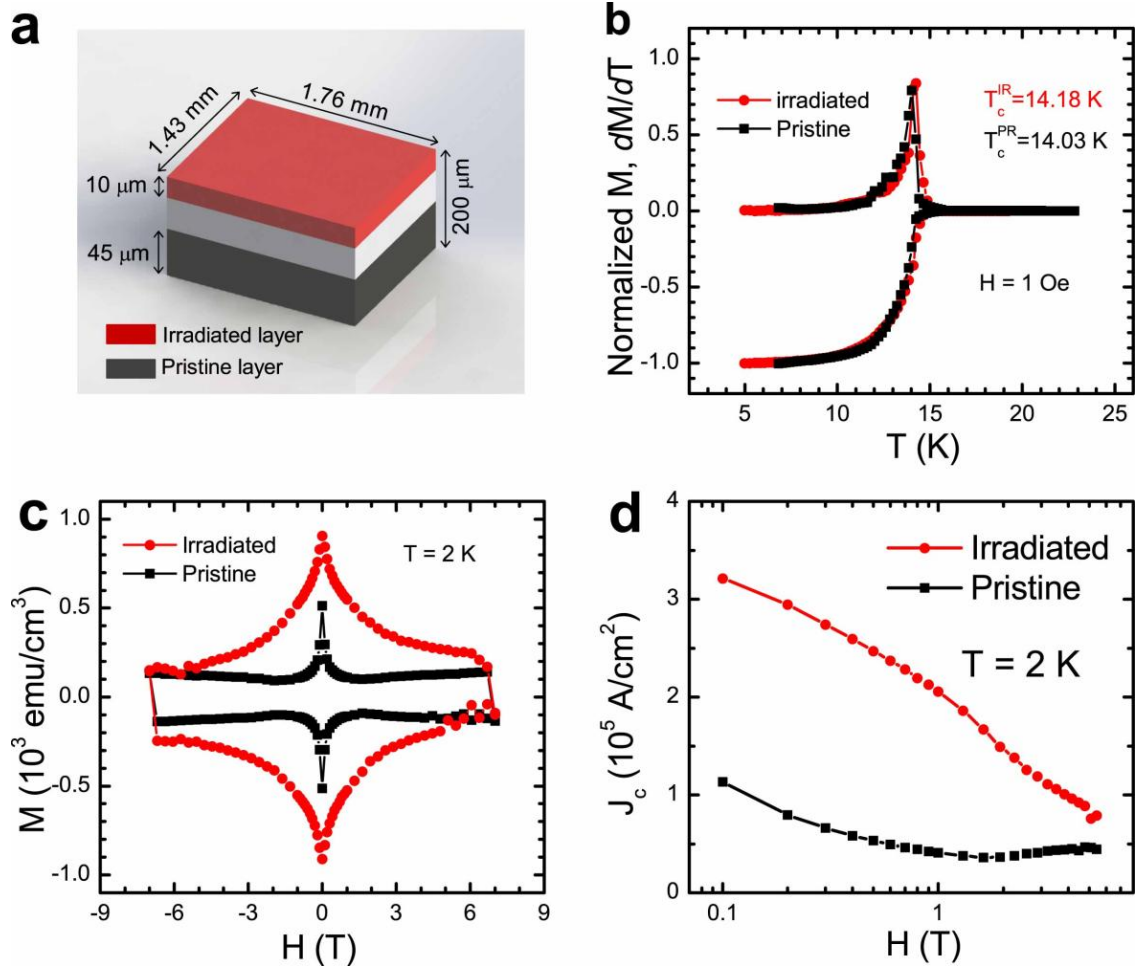

**Fig. S1   Magnetization and critical current density**

**a**     Schematic drawing of the irradiated Fe(Se, Te) sample of dimension 1.43 * 1.76 mm$^2$ measured with STM. Indicated are the irradiated and the pristine region used in the measurements presented in (b), (c), and (d): After performing STM measurements on the top (irradiated) surface, magnetization measurements were performed on  a thin layer of the same top surface (irradiated) and on a layer of bottom surface (pristine).

**b**     Normalized magnetization and its derivative with respect to temperature as a function of temperature of the irradiated and the pristine region of the sample.

**c**     Magnetization as a function of applied field of the irradiated and pristine region of the sample.

**d**     Critical current density as a function of field for the irradiated and the pristine region of the sample.

**Fig. S2**

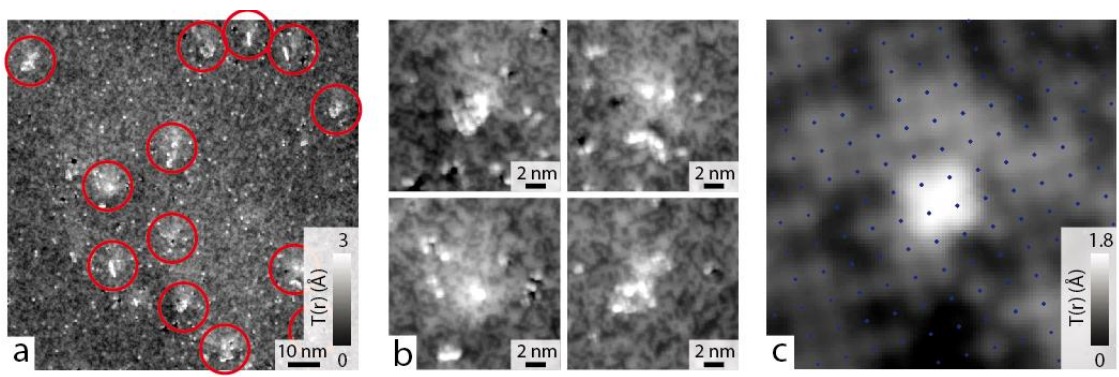

**Fig. S2   Columnar and point defects in more detail**

**a**     Constant current image as also shown in Fig. 3a, with all columnar defects indicated by red circles.

**b**     Enlargements of four of the columnar defects in panel (a) using the same color bar. At the core of the columnar defect, no atomic lattice can be discerned; instead, an amorphous region with a larger corrugation is seen.

**c**     Typical point defect. The position of the Se and Te atoms is indicated with blue dots for clarity (atomic spacing is 3.9Å). The point defect is centered in between the Se/Te atoms, corresponding to the Fe position in the layer below.

**Fig. S3**

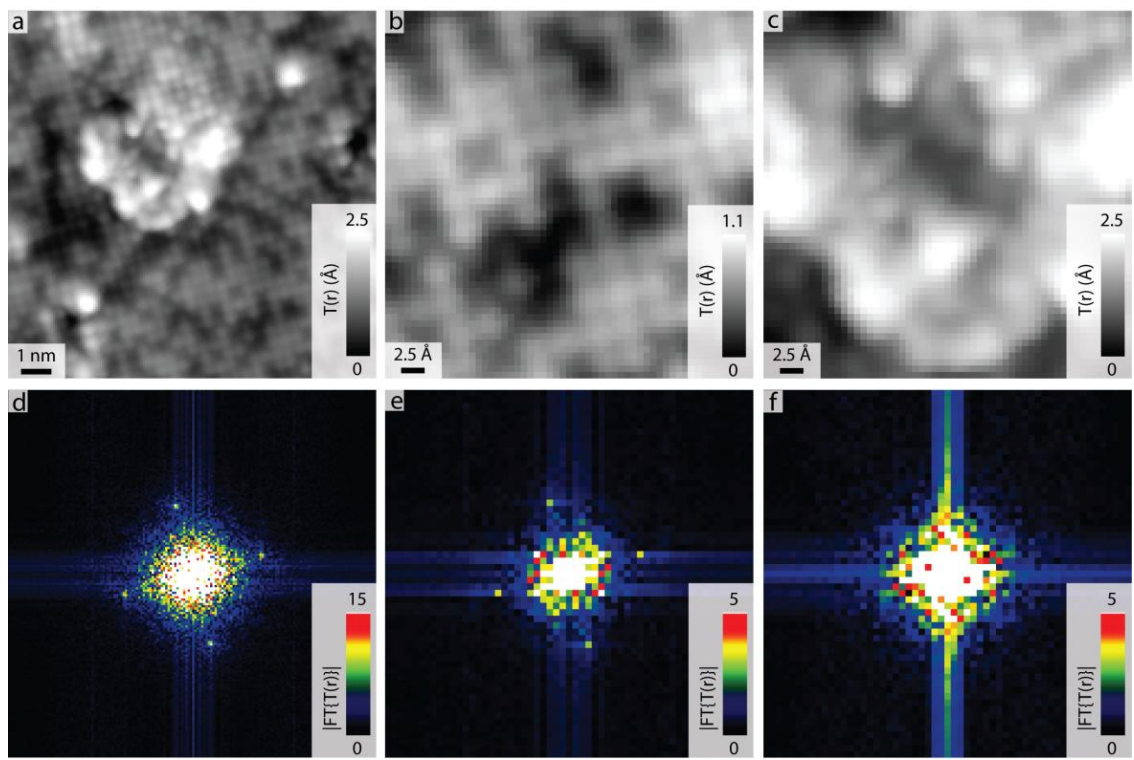

**Fig. S3   Crystal structure of the columnar defects**

**a**     High resolution constant current topography of a columnar defect and the surrounding crystal lattice.

**b**     High resolution topography showing only the crystal lattice.

**c**     High resolution topography showing only the columnar defect from **a**. No crystal lattice can be discerned inside the defect region.

**d**     Fourier transform of topograph shown in **a**: Bragg peaks of the crystal lattice are clearly visible.

**e**     Fourier transform of crystal lattice shown in **b**: As expected the Bragg peaks are clearly discernible.

**f**     Fourier transform of columnar defect shown in **c**: No Bragg peaks are visible, and instead long wavelength signals dominate the Fourier transform as expected in the case of an amorphous structure lacking the short wavelength periodicity of the crystal lattice.

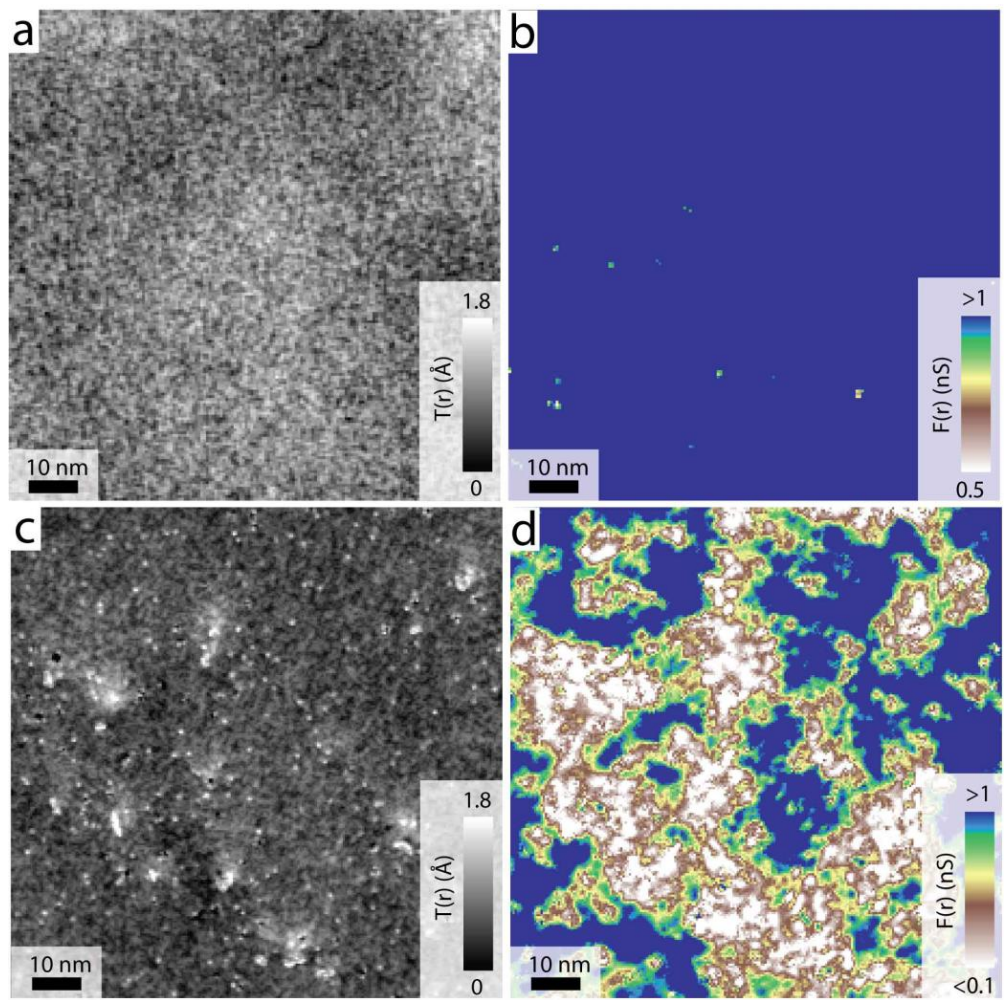

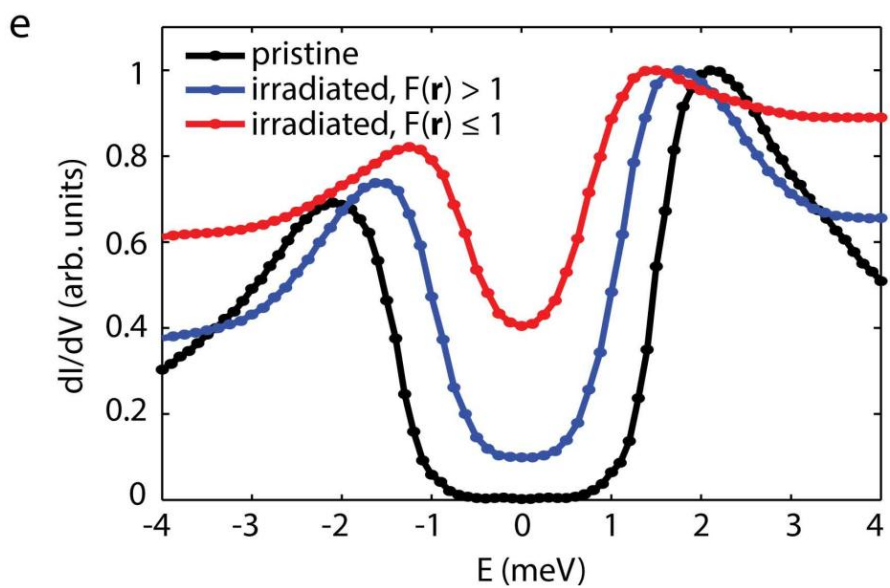

**Fig. S4  F(r) for pristine and irradiated Fe(Se,Te) compared**

**a**    Constant current topography of the pristine Fe(Se, Te) sample of dimension 100 * 100 nm² recorded at 270 mK and 0 T. No intrinsic defects are visible.

**b**    F(r) for the pristine sample of the same field of view as (a).

**c**    Constant current topograph of the irradiated Fe(Se, Te) sample of dimension 100 * 100 nm² recorded at 270 mK and 0 T. Numerous point- and several columnar defects are visible.

**d**    F(r) for the irradiated sample of the same field of view as (c).

**e**    Comparison of the average spectra at 270 mK of the pristine sample, the region in the irradiated sample which is quasi defect-free (F(r) > 1), and the region which contains the point and columnar defects (F(r) ≤ 1).

**Fig. S5**

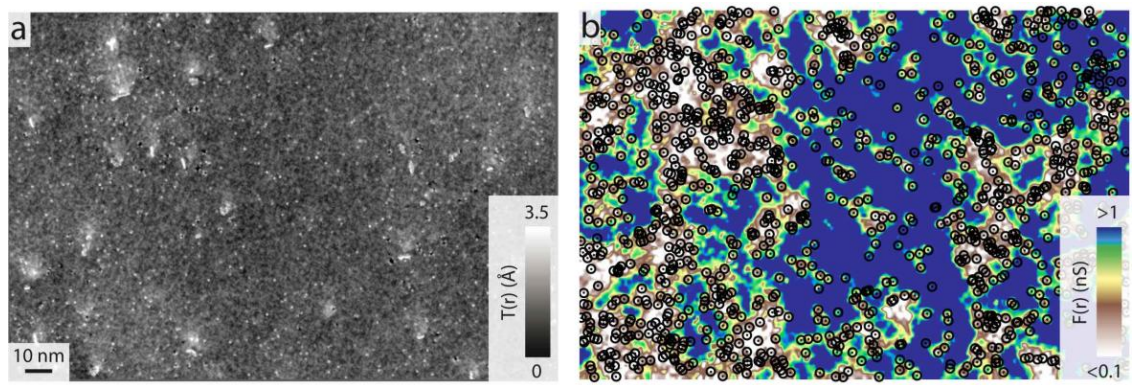

**Fig. S5   Effect of point defects on superconductivity**

**a**    Constant current topography of irradiated Fe(Se, Te).

**b**    F(r) map simultaneously recorded with (a). The point-like defects are indicated with small black circles.

**Fig. S6**

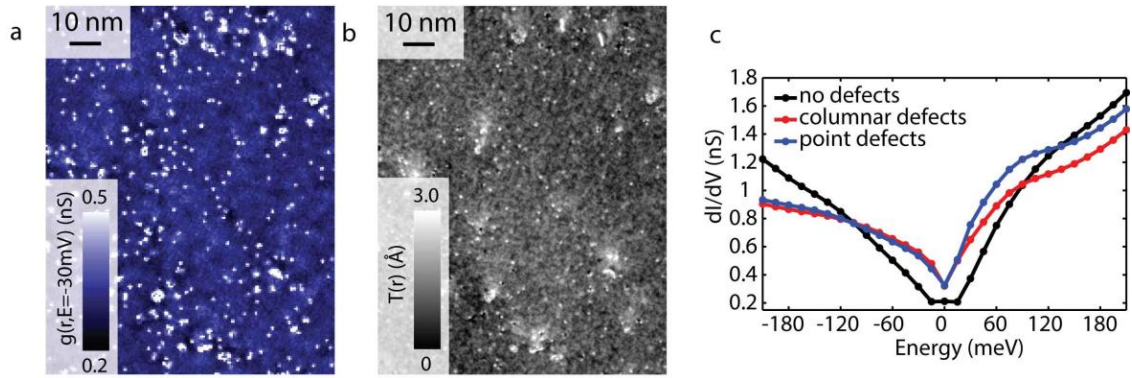

**Fig. S6   High energy (normal state) characteristics of damage**

**a**      Differential conductance map layer at -30 mV. Point and columnar defects are clearly visible at this energy as a comparison with the topograph in (b) shows.

**b**      High resolution constant current topography of same field of view as (a).

**c**      Average spectra at 1.2 K for regions containing no defects, point defects, and columnar defects in the field of view shown in (a) and (b).

**Fig. S7**

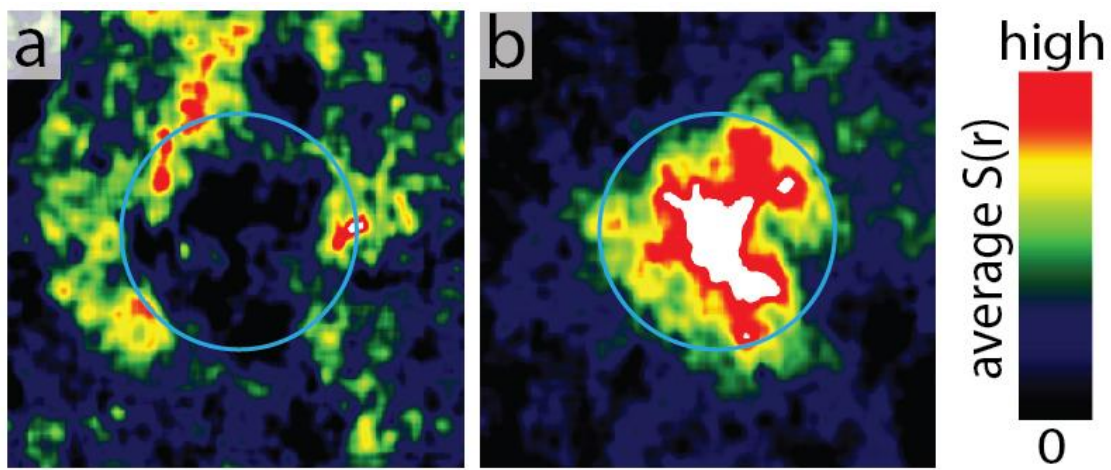

**Fig. S7   Vortex halos at columnar defects**

**a**    Average S(r) of the columnar defect locations, reproduced from main text Fig.  3c.

**b**    Average S(r) of the clearly observable, 'traditional', vortices in the field of view of main text Fig. 3b. Panel a and b are both 20x20nm². The circles in both panels are of identical size and show that the hole inside the halo corresponds very well to the core of the average vortex.